\documentclass[aps,prd,notitlepage,showpacs,nofootinbib,superscriptaddress]{revtex4-1}

\usepackage{graphicx}
\usepackage[utf8]{inputenc} 
\usepackage{subfigure}
\usepackage{amsmath}
\usepackage{amssymb}
\usepackage{braket}
\usepackage{float}
\usepackage{comment}
\usepackage{slashed}
\usepackage[normalem]{ulem}
\usepackage{adjustbox}

\usepackage{caption}
\usepackage{subcaption}
\usepackage{amsmath}
\usepackage{amssymb}

\usepackage[usenames,dvipsnames]{color}
\usepackage[colorinlistoftodos]{todonotes}
\usepackage[colorlinks=true,citecolor=darkred,urlcolor=darkred, pdfborder={0 0 0}]{hyperref}
\usepackage[normalem]{ulem}

\definecolor{darkred}{rgb}{0.6,0,0}
\definecolor{darkbrown}{rgb}{0.39,0.26,0.12}
\usepackage[colorinlistoftodos]{todonotes}

\definecolor{linkcolor}{rgb}{0,0,0.5}
 
\newcommand {\ignore}[1]{}

\def\be{\begin{equation}}
\def\ee{\end{equation}}

\def\gsim{\raise0.3ex\hbox{$\;>$\kern-0.75em\raise-1.1ex\hbox{$\sim\;$}}}
\def\lsim{\raise0.3ex\hbox{$\;<$\kern-0.75em\raise-1.1ex\hbox{$\sim\;$}}}

\usepackage{soul}

\definecolor{mightnightblue}{RGB}{25,25,112}

\definecolor{brown}{rgb}{0.59, 0.29, 0.0}

\def\21{$\mathrm{SU(2)_L \otimes U(1)_Y}$}

\begin{document}

\bibliographystyle{unsrt}   

\title{
\color{BrickRed} Probing Inelastic Dark Matter via Cosmic-Ray Upscattering in NGC 1068} 

\author{Eung Jin Chun}\email{ejchun@kias.re.kr}
\affiliation{Korea Institute for Advanced Study, Seoul 02455, Korea}

\author{Sanjoy Mandal}\email{smandal@kias.re.kr}
\affiliation{Korea Institute for Advanced Study, Seoul 02455, Korea}

\author{Abhishek Roy}\email{abhishek@sogang.ac.kr}
\affiliation{Center for Quantum Spacetime, Sogang University, 35 Baekbeom-ro, Mapo-gu, Seoul, 121-742, South Korea}
\affiliation{Department of Physics, Sogang University, 35 Baekbeom-ro, Mapo-gu, Seoul, 121-742, South Korea}

\begin{abstract}
We study constraints on sub-GeV inelastic dark matter (iDM) from cosmic-ray (CR) cooling in the active galactic nucleus (AGN) NGC 1068. In dense dark matter~(DM) spikes surrounding supermassive black holes, high-energy CR protons can efficiently lose energy through scatterings with dark matter particles. We consider a minimal vector-portal iDM framework and consistently include both elastic and deep inelastic scattering (DIS) contributions to the CR energy-loss rate. We find that DIS processes dominate at high momentum transfer and substantially enhance the DM-induced cooling effect. By requiring the resulting cooling timescale to remain compatible with the observed Standard Model cooling in NGC 1068, we derive constraints on the iDM parameter space. Our results demonstrate that AGN cosmic-ray cooling probes previously unexplored regions of sub-GeV iDM parameter space inaccessible to current direct-detection experiments.

\end{abstract}

\maketitle


\section{Introduction}
The presence of dark matter~(DM) in galaxies and galaxy clusters is well established by diverse astrophysical and cosmological observations, yet its fundamental particle nature remains unknown~\cite{Bertone:2004pz,Cirelli:2024ssz,Bergstrom:2000pn}. Weakly interacting massive particles~(WIMPs) are compelling dark matter candidates~\cite{Lee:1977ua,Jungman:1995df} because of their weak-scale interactions, natural thermal production in the early universe, and strong motivation from extensions of the Standard Model. However, the absence of definitive detection has shifted focus toward a wider class of candidates, particularly in the sub-GeV mass range. Yet conventional direct detection experiments rapidly lose sensitivity in this regime due to kinematic threshold limitations~\cite{Essig:2012yx,Aprile:2012zx,PandaX-II:2017hlx,Dolan:2017xbu,Bell:2021ihi,Elor:2021swj,DAMIC-M:2023gxo,SENSEI:2023zdf,DeMarchi:2024riu}. 
Several approaches have been proposed to extend direct detection sensitivity to sub-GeV dark matter, including boosted components produced through gravitational effects or scatterings with protons, electrons, or neutrinos in astrophysical environments~\cite{Baushev:2012dm, Besla:2019xbx, Herrera:2021puj, Herrera:2023fpq, Smith-Orlik:2023kyl, Bringmann:2018cvk, Ema:2018bih, Alvey:2019zaa, Wang:2021jic, Agashe:2014yua, Kim:2016zjx, Das:2021lcr, Cappiello:2022exa, Arguelles:2022fqq,Lin:2022dbl}.
\par There exists a class of sub-GeV dark matter models in which present-day annihilation is suppressed due to non-standard evolution mechanisms, such as number-changing $3\to 2$ processes in the dark sector~\cite{Hochberg:2014dra,Fitzpatrick:2020vba,Ho:2022erb}, asymmetric dark matter~\cite{Bhattacherjee:2013jca,Izaguirre:2015yja,Ho:2022tbw}, and freeze-in production~\cite{Frangipane:2021rtf,Bhattiprolu:2022sdd}. As a result, this class of light dark matter can be probed only through direct detection and accelerator experiments. 

A particularly important framework in this context is inelastic dark matter~(iDM) \cite{Tucker-Smith:2001myb,Tucker-Smith:2004mxa,Chang:2008gd} which was initially introduced to account for the anomalous observations reported by the DAMA collaboration~\cite{Bernabei:2013xsa}. Since then, iDM has  developed into a well-motivated paradigm for sub-GeV thermal dark matter interacting via a vector mediator~\cite{Jordan:2018gcd,Berlin:2018pwi,Batell:2021ooj,Mongillo:2023hbs,Izaguirre:2015zva}. In this scenario, the dark matter spectrum consists of states with a small mass splitting, leading to distinctive experimental signatures. A defining feature of these models is the presence of off-diagonal interactions between the ground state, $\chi_1$, and a slightly heavier excited state, $\chi_2$ with $m_{\chi_2}>m_{\chi_1}$. This structure leads to kinematic suppression of dark matter–nucleon scattering rates in present-day direct detection experiments. As a result, iDM models are only weakly constrained by direct detection experiments, particularly when the dark matter mass lies in the sub-GeV range and the mass splitting is relatively large~\cite{Tucker-Smith:2001myb, Schwetz:2011xm,Bramante:2016rdh, Baryakhtar:2020rwy, Song:2021yar, Bell:2021xff, Herrera:2023fpq, Eby:2023wem, Chatterjee:2022gbo, Kang:2024kec,Emken:2021vmf, Herrera:2023fpq, Garcia:2024uwf, Essig:2022dfa, Yun:2023huf, Li:2022acp, Bell:2022dbf, Gu:2022vgb}.
This motivates the exploration of extreme astrophysical environments that may provide enhanced sensitivity to the non-annihilating sub-GeV DM scenarios.
\par Active galactic nuclei~(AGNs) are powerful environments that efficiently accelerate cosmic rays~(CRs), where accretion onto a supermassive black hole~(SMBH) drives jets, shocks, and turbulent magnetic fields that boost particles to ultra-high energies~\cite{Rieger:2022qhs}. In these environments, accelerated CRs—primarily protons—can interact with ambient matter ($pp$ interactions) or radiation fields ($p\gamma$ interactions), producing secondary particles such as pions~\cite{Murase:2022feu}. The decay of these pions leads to high-energy gamma rays and neutrinos, making AGNs promising multimessenger sources. The
CR protons also cool through various other SM processes like synchrotron, inverse Compton, Bethe-Heitler pair production processes and adiabatic losses. For AGNs such as, NGC 1068~\cite{IceCube:2022der,Ajello:2023hkh} and TXS 0506+056~\cite{IceCube:2018dnn,IceCube:2018cha,MAGIC:2018sak,Keivani:2018rnh,Padovani:2019xcv}, detailed multimessenger observations and modeling are available, allowing us to assess the energy dependence of cooling times.  Also SMBH at the Galactic Center~(GC) can enhance the surrounding DM density, forming a steep spike over the cosmic time as its gravity pulls in and compresses the halo beyond standard predictions~\cite{Gondolo:1999ef,Ullio:2001fb,Lacroix:2016qpq}.
As CRs propagate through these dense DM environments, energy transfer from CRs to DM can induce significant cooling effects, making AGNs promising laboratories for probing DM--proton and DM--electron interactions in the sub-GeV regime~\cite{Herrera:2023nww, Gustafson:2025dff,Li:2025zwg,Gustafson:2024aom,Mishra:2025juk}.

In this work, we focus primarily on NGC~1068, which provides stronger and more robust sensitivity to iDM compared to blazar systems such as TXS~0506+056. This is mainly due to its dense, proton-dominated environment and the presence of ultra-high-energy CR protons spanning energies from approximately $10^2$ to $10^7~\mathrm{GeV}$. Such a broad CR spectrum enables efficient energy transfer to DM particles and substantially enhances CR cooling through DM upscattering. By contrast, TXS~0506+056 is largely characterised by CR electrons with a comparatively narrower energy range, limiting the maximum transferable energy and suppressing the overall cooling effect. Consequently, proton-induced cooling in NGC~1068 leads to significantly stronger and more reliable constraints on iDM parameter space.

We consider a minimal vector-portal iDM framework in which DM interacts with protons through off-diagonal couplings. Previous studies of sub-GeV DM scattering have primarily focused on elastic processes described by dipole form factors. However, Ref.~\cite{Li:2025zwg} demonstrated that this treatment becomes inadequate at high proton energies, where the internal structure of the proton can no longer be neglected and deep inelastic scattering~(DIS) channels become important. Therefore, analyses restricted to low momentum transfer fail to provide reliable constraints in the high-energy CR regime relevant for AGNs. Motivated by this observation, we consistently include both elastic and DIS contributions to CR--DM scattering when evaluating the CR cooling timescale.

Finally, throughout this work we assume that $\chi_1$ constitutes the present-day DM component, while remaining agnostic about the precise mechanism responsible for generating the observed relic abundance. Viable scenarios include co-annihilation or co-scattering mechanisms involving heavier dark-sector states~\cite{Belanger:2025wjh,Griest:1990kh,Garny:2017rxs}, as well as non-standard cosmological histories such as early matter domination or kination-dominated eras~\cite{Gelmini:2006pw,Bernal:2024yhu,Roy:2025moo,Belanger:2024yoj,Ferreira:1997hj,Joyce:1996cp,DEramo:2017gpl}. Since these possibilities can naturally reproduce the observed DM abundance while leaving the present phenomenology largely unchanged, our analysis is intended to remain broadly applicable to a wide class of iDM realisation.

\section{BSM Cooling of CRs in AGN}
Multimessenger data from NGC 1068 enable us to estimate the energy-dependent cooling times of CR protons, dominated by Standard Model processes such as $pp$ and $p\gamma$ interactions, as well as Bethe–Heitler pair production~\cite{Murase:2019vdl, Fiorillo:2024akm, Fiorillo:2025ehn}~\footnote{Specifically, for proton energies $T_p \lesssim 10^4$ GeV, cooling is dominated by $pp$ interactions with the ambient gas. In the intermediate range $T_p \sim 10^4\text{--}10^6$ GeV, Bethe–Heitler pair production becomes significant, while for $T_p \gtrsim 10^6$ GeV, energy losses are primarily driven by $p\gamma$ interactions~\cite{Murase:2022dog}.}. For NGC 1068, detailed multimessenger observations and modeling are available, allowing us to evaluate the energy dependence of the cooling times. The behaviors of the cooling timescales with the CR energy is shown by black solid line in Fig.~\ref{fig:cooling}.
\par Furthermore, it has been suggested that the adiabatic growth of a SMBH can give rise to a DM spike in its surrounding region~\cite{Gondolo:1999ef,Ullio:2001fb,Lacroix:2016qpq}. Starting from an initial power-law profile $\rho(r) = \rho_{0}\left( r/r_{0} \right)^{-\gamma}$, the distribution evolves into:
\begin{align}
	\rho_{\rm sp}(r) = \rho_{R} \, g_{\gamma}(r)\, \bigg(\frac{R_{\rm sp}}{r}\bigg)^{\gamma_{\rm sp}}\;,
\end{align}
where  $R_{\mathrm{sp}} = \alpha_{\gamma} r_0 \left( M_{\mathrm{BH}}/(\rho_{0}r_0^{3}) \right) ^{\frac{1}{3-\gamma}}$ is the size of the
spike, and $\gamma_{\mathrm{sp}} = (9-2\gamma)/(4-\gamma)$. Further, $g_{\gamma}(r) \approx (1-\frac{4R_{\rm S}}{r})$, while $\rho_{\rm R}$ is a normalization factor, chosen to match the density profile outside of the spike, $\rho_R=\rho_{0}\, (R_{\rm 
sp}/r_0)^{-\gamma}$. This density profile vanishes at $4 R_{\rm S}$, which represents a conservative approximation. We consider that the initial DM distribution follows an NFW profile \cite{Navarro:1996gj,Navarro:1995iw}, with $\gamma=1$, resulting in a spike with $\gamma_{\rm sp}=7/3$ and $\alpha_{\gamma}= 0.122$. For the scale radius and SMBH mass of NGC 1068, we take $r_0=10$ kpc and $10^{7}M_{\odot}$, respectively. Note that if DM undergoes self-annihilation, the density in the innermost region of the spike is limited and saturates at $\rho_{\rm sat} = \frac{m_{\rm DM}}{\langle \sigma v \rangle\, t_{\rm BH}}$ ,
where $\langle \sigma v \rangle$ is the velocity-averaged annihilation cross section and $t_{\rm BH}$ denotes the age of the central SMBH which we considered to be $10^{10}$ yr for NGC 1068. We find that the average density of asymmetric or non-annihilating DM particles in the corona of
NGC 1068 is $\left<\rho_{\rm DM}\right>\sim 5\times 10^{18}\text{ GeV}/\text{cm}^3$, indicating that the DM density can be extremely high in the region where
high-energy CR protons are produced.
\par As CRs pass through this DM spike, they scatter off DM particles.   When the scattering cross section is sufficiently large, this leads to efficient CR cooling, which can in turn  modify the high-energy neutrino and gamma-ray fluxes emitted by these sources. As a result, observations of neutrinos and gamma rays offer powerful constraints on CR cooling driven by DM–proton interactions~\cite{Herrera:2023nww}. The BSM cooling timescale due to DM-proton scattering  is given by~\cite{Ambrosone:2022mvk}
\begin{equation}
    t_{\chi_1 p}=\left[-\frac{1}{E}\left(\frac{\mathrm{~d} E}{\mathrm{~d} t}\right)_{\chi_1 p}\right]^{-1},
\label{eq:timescale}
\end{equation}
where $E$ is the CR energy and $\dot{E}_{\chi_1 p} \equiv \left(\frac{\mathrm{~d} E}{\mathrm{~d} t}\right)_{\chi_1 p}$ is the energy loss rate due to CR–DM collisions, defined as
\begin{equation} \label{eq:InelELoss}
\dot{E}_{\chi_1 p}= -\frac{\langle \rho_{\chi_1}\rangle}{{m_{\chi_1}}}\,\int_{T^{\rm min}_{\rm DM}}^{T^{\rm max}_{\rm DM}} dT_{\rm DM}\, (T_{\rm DM} + \delta_{\rm DM}) \frac{d\sigma_{\chi_1 \, p \rightarrow \chi_2 X}}{dT_{\rm DM}},
\end{equation}
where $\langle \rho_{\chi_1} \rangle$ is the average density of DM particles in the region of CR production. $d\sigma_{\chi_1 \, p \rightarrow \chi_2 X}/dT_{\rm DM}$ is the differential DM-proton cross section, $T^{\rm max}_{\rm DM}~(T^{\rm min}_{\rm DM})$ is the maximal~(minimal) allowed value for the upscattered DM kinetic energy $T_{\rm DM}$ in a collision with proton with kinetic energy $T_p = E_p-m_p$. The functional form of the differential distributions $d\sigma_{\chi_1 \, p \rightarrow \chi_2 X}/dT_{\rm DM}$ differs fundamentally between elastic ($X=p$)~\footnote{Traditionally, the process $\chi_1\, p\to\chi_2\, p$ is classified as inelastic scattering since the final-state masses differ from the initial-state masses. However, in this work we refer to it as elastic scattering to distinguish it from DIS.} and deep inelastic scattering (DIS; $X=\text{quarks}$) processes. The kinematics energy transfer $T_{\rm DM}^{\rm min/max}$ also differs depending on the type of processes.
\par The functional form of the differential distribution for both elastic scattering and DIS depends on the underlying particle physics model. We consider a simplified dark sector consisting of a vector mediator $Z'$, a stable fermion DM state $\chi_1$ with mass $m_{\chi_1}$, and a heavier fermion $\chi_2$ with mass $m_{\chi_1}+\delta_{\rm DM}$. The interaction between the SM and the dark sector is taken to be of the form~\cite{Tucker-Smith:2001myb,Tucker-Smith:2004mxa,Chang:2008gd}
\begin{align}
\mathcal{L}=\Big[g_\chi\bar{\chi}_1\gamma^\mu\chi_2 Z'_{\mu}  + \text{h.c} \Big] + \sum_{f}g_f\bar{f}\gamma^\mu f Z'_{\mu},
\end{align}
where $g_{\chi}$ and $g_f$ are the coupling constants of the $Z'$ to the dark sector and SM fermions, respectively.  We assume that $Z'$ couples equally to protons and neutrons, $g_p=g_n=g_q$, and set other SM couplings to zero. When CR protons transfer only a small momentum to DM, such that  $Q^2 \ll 1~\mathrm{GeV}^2$ (where $Q^2=-q^2 = -(p-p')^2$, with $p$ and $p'$ denoting the four-momenta of the initial- and final-state protons, respectively), the proton can effectively be treated as a point-like particle and the effects of its internal structure remain negligible. As the momentum transfer increases, however, the nucleon substructure becomes increasingly important, for example through its charge and magnetization distributions.  In this region, the proton-$Z'$ interaction can be defined by the electromagnetic form factors $F_1(Q^2)$ and $F_2(Q^2)$~\cite{Perdrisat:2006hj},
\begin{equation}
  \bra{N(p^{\prime})} \sum_{q} g_{q} \bar{q} \gamma^{\mu} q Z_{\mu}^{\prime} \ket{N(p)} = g_p Z_{\mu}^{\prime} \bar{N} \Gamma^{\mu} N, \,\,\mathrm{with} \quad   \Gamma^{\mu}=\gamma^{\mu} F_{1}\left(Q^{2}\right)+\frac{i}{2 m_{p}} \sigma^{\mu \nu} q_{\nu} F_{2}\left(Q^{2}\right),
  \label{eq:effint}
\end{equation}
At low $Q^2$, the Pauli form factor $F_2$ is generally suppressed by $\mathcal{O}(q/m_p)$, whereas the Dirac form factor $F_1$ can be accurately approximated using the dipole form: $F_1(Q^2)=(1+Q^2/\Lambda^2)^{-2}$ with $Q^2=2m_{\chi_1}T_{\rm DM}-\delta_{\rm DM}^2$ and $\Lambda=770$~MeV. At higher momentum transfer, $F_2$ becomes non-negligible and the dipole form
fails. Hence one need to therefore retain both form factors~\footnote{It is often convenient to parameterize the form factors $F_1$ and $F_2$ in terms of the Sachs form factors $G_E^p$ and $G_M^p$, which independently characterize the electric and magnetic distributions of the proton~\cite{Sachs:1962zzc,Kelly:2004hm}.}. As a result, we obtain the differential cross section
of DM-proton elastic scattering as,
\begin{align}
\frac{\mathrm{d}\sigma_{\chi p}}{\mathrm{d} T_{\rm DM}}= \frac{\overline{\left | \mathcal{M}  \right |^{2}}  }{32\pi m_{\chi}T_p\left ( T_p+2m_{p} \right )} 
&= \frac{ g_{p}^{2} g_{\chi}^{2}}{8\pi T_p\left ( T_p+2m_{p} \right )\left(Q^{2}+m_{Z^{\prime}}^{2}\right)^{2}}\nonumber\\
&\times [(F_{1}\left(Q^{2}\right) 
 +F_{2}\left(Q^{2}\right)  )^{2}\mathcal{K}_{1}+(1+\tau) F_{2}\left(Q^{2}\right)^{2}\mathcal{K}_{2}],
 \label{eq:diff_elastic}
\end{align}
where $\tau=Q^2/(4m_p^2)$ and the function $\mathcal{K}_{1,2}$ are given as,
\begin{align}
&\mathcal{K}_1 = -2 m_{\chi_1}^2  T_{\rm DM} - T_{\rm DM}(2m_p^2+\delta_{\rm DM}^2) + m_{\chi_1} (4E_p^2-4E_p T_{\rm DM}+2T_{\rm DM}^2 - 4E_p \delta_{\rm DM} + \delta_{\rm DM}^2),\\
& \mathcal{K}_2=m_{\chi_1} (2E_p-\delta_{\rm DM}) (2(E_p-T_{\rm DM})-\delta_{\rm DM})+\frac{1}{2}T_{\rm DM}(\delta_{\rm DM}^2-4m_p^2).
\end{align}
This expression allows one to evaluate the energy-loss rate via Eq.~\ref{eq:timescale}, which in turn determines the CR cooling timescale where the maximum and minimum upscattered DM kinetic energy are given as,
\begin{align}
T_{\rm DM}^{\rm min/max}=\frac{ E_p+m_{\chi_1}}{2s} \Big[s+\left(m_{\chi_1}+\delta_{\rm DM}\right)^2- m_p^2 \pm \frac{\sqrt{E_p^2-m_p^2}}{E_p+m_{\chi_1}}\lambda^{\frac{1}{2}}\left(s, m_p^2,\left(m_{\chi_1}+\delta_{\rm DM}\right)^2\right)\Big ] -\left(m_{\chi_1}+\delta_{\rm DM}\right),
\end{align}
where $s=m_p^2+m_{\chi_1}^2+2\, m_{\chi_1} E_p$.
\par At sufficiently high proton energies, DM-proton scattering is dominated by DIS. Within the parton model, the nucleon is treated as a collection of quasi-free, spin-$1/2$, point-like partons, and the DM–proton DIS process is approximated as an incoherent sum of elastic DM–parton scatterings, $\chi_1(k)+q(xP)\to \chi_2(k')+q(p')$, where $x$ is the momentum fraction carried by the struck parton. In the rest frame of DM, the differential DIS cross section reads,
\begin{equation}
\begin{aligned}
    \frac{\mathrm{d} \sigma_{\chi p}}{\mathrm{d} T_{\rm DM }} &= \sum_{q}\int_{x_{\min }}^{1} dx\, f_q\left(x, Q^{2}\right) \frac{d\hat{\sigma}}{dT_{\rm DM}}, 
\end{aligned}
\end{equation}
where the partonic cross section is 
\begin{align}
\frac{d\hat{\sigma}}{dT_{\rm DM}}=\frac{m_{\chi_1}}{8\pi\lambda(\hat{s},m_q^2,m_{\chi_1}^2)}  \frac{2g_q^2 g_{\chi}^2}{(Q^2+M_{Z'}^2)^2}\Big[(Q^2+\delta_{\rm DM}^2) & (Q^2-2m_q^2-4xE_p m_{\chi_1})\nonumber \\
&-2m_{\chi_1}^2\left(Q^2+4xE_p(\delta_{\rm DM}-xE_p)\right)\Big],
\label{eq:diff_DIS}
\end{align}
where $Q^2=-\hat{t}=-\delta_{\rm DM}^2+2m_{\chi_1}T_{\rm DM}$ and $\hat{s}=m_q^2+m_{\chi_1}^2+2\,x\, m_{\chi_1} E_p$. The parton distribution function $f_{q}(x,Q^2)$ describes the probability of finding a parton carrying a momentum fraction $x$ of the proton. We only take the partonic contributions from up and down quarks. For this we considered the CT18nnlo set implemented in the LHAPDF package~\cite{Buckley:2014ana}. The lower limit on $x$ is fixed by the threshold condition $\hat{s}>(m_q+m_{\chi_1}+\delta_{\rm DM})^2$,
\begin{align}
x_{\rm min}=\frac{m_q+\delta_{\rm DM}}{E_p}+\frac{\delta_{\rm DM}}{2E_p\, m_{\chi_1}}\left(\delta_{\rm DM}+2m_q\right).
\end{align}
The maximum and minimum upscattered DM kinetic energy are given as
\begin{align}
T_{\rm DM}^{\rm min/max}=\frac{x\, E_p+m_{\chi_1}}{2\hat{s}} \Big[\hat{s}+\left(m_{\chi_1}+\delta_{\rm DM}\right)^2-m_q^2 \pm \frac{x\sqrt{E_p^2-m_p^2}}{x\, E_p+m_{\chi_1}}\lambda^{\frac{1}{2}}\left(\hat{s}, m_q^2,\left(m_{\chi_1}+\delta_{\rm DM}\right)^2\right)\Big ] -\left(m_{\chi_1}+\delta_{\rm DM}\right).
\end{align}
Note that the maximum momentum transfer in both the elastic scattering and DIS cases is given by
$Q^2_{\rm max} = -\delta_{\rm DM}^2 + 2 m_{\chi_1} T_{\rm DM}^{\rm max} \, .$ For elastic scattering, the maximum DM kinetic energy satisfies $T_{\rm DM}^{\rm max} < T_p \, .$
In contrast, for DIS processes, the hadronization of the final state relaxes the momentum-transfer constraint, allowing nearly the full proton energy to be transferred to the DM particle, i.e.
$T_{\rm DM}^{\rm max} \approx T_p \, .$ Consequently, one obtains
$Q^2_{\rm max,\,DIS} > Q^2_{\rm max,\,ES} \, ,$
and this enhancement plays a crucial role in making the DIS contribution significantly larger than the elastic scattering contribution.
\par In our analysis, we include both the elastic and DIS contributions to the cooling timescale. To derive constraints on the iDM model, we impose the CR cooling condition $t_{\chi_1 p} \leq C\, t_{\rm SM}$ to exclude regions of parameter space in which the cooling proceeds too rapidly. The factor $C$ is model dependent parameter and for NGC 1068 it was estimated to vary in the range $0.1\leq C\leq 1$~\cite{Herrera:2023nww}. For our calculations, we adopt a conservative value of $C=0.1$.
\section{Results and discussion}
To compare the contributions of elastic scattering and DIS to the CR cooling rate, we evaluate the energy loss associated with each process focusing on sub-GeV DM with negligible present day annihilation. 
\begin{figure}[]
\centering
\mbox{
\subfigure[]{\includegraphics[width=0.5\textwidth]{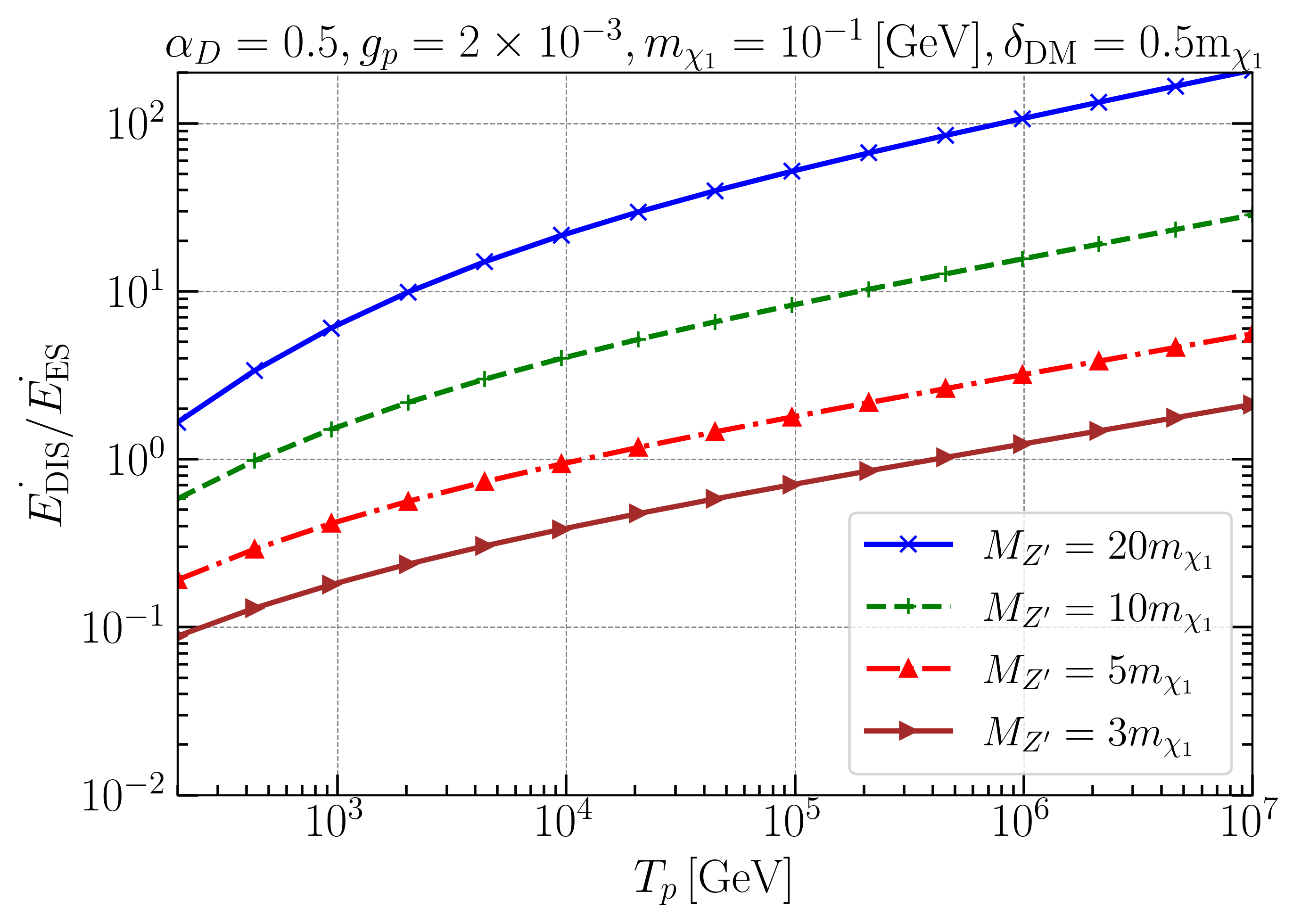}\label{Fig:BP_1a}}
\subfigure[]{\includegraphics[width=0.5\textwidth]{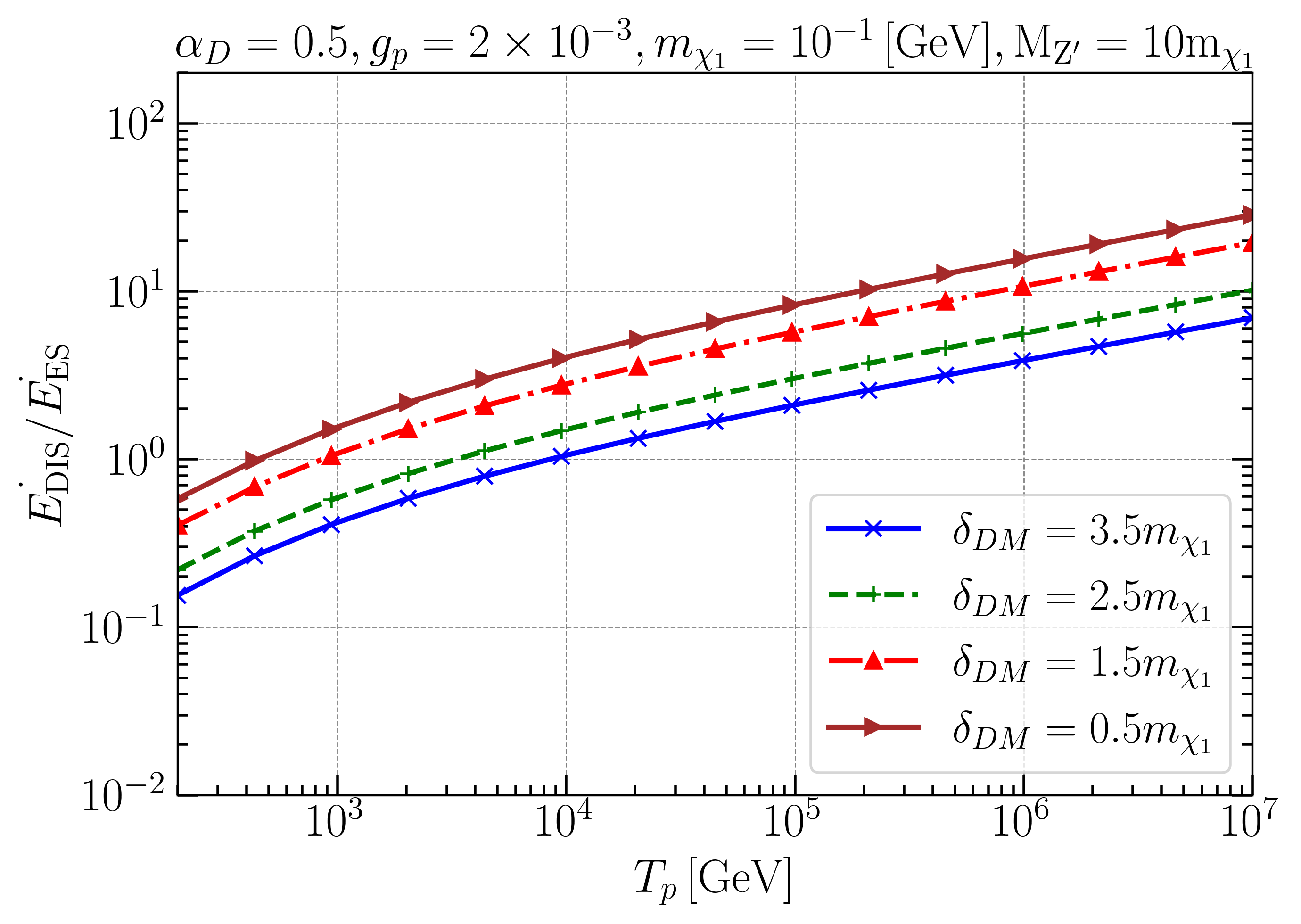}\label{Fig:BP_1b}}
}
\caption{Ratio of DIS to elastic scattering (ES) energy loss rate $\dot{E}_{\rm DIS}/\dot{E}_{\rm ES}$ as a function of proton kinetic energy in CRs for fixed DM mass $m_{\chi_1}=0.1$~GeV. The left and right panel shows the impact on this ratio due to mediator mass~($M_{Z'}$) and mass splitting~($\delta_{\rm DM}$), respectively.}
\label{Fig:BP_Plot1}
\end{figure}
Fig.~\ref{Fig:BP_Plot1} shows the ratio of DIS to elastic scattering (ES) energy loss rate $\dot{E}_{\rm DIS}/\dot{E}_{\rm ES}$ as function of proton kinetic energy in CRs where we fixed the DM mass as $m_{\chi_1}=0.1$~GeV. The left panel shows the dependence on the mediator mass $M_{Z'}$ at fixed mass splitting $\delta_{\rm DM}=0.5\, m_{\chi_1}$, while the right panel shows the dependence on  the mass splitting at fixed mediator mass $M_{Z'}=10\, m_{\chi_1}$. It is evident that both $M_{Z'}$ and $\delta_{\rm DM}$ play important roles in determining the scattering cross section and, consequently, the CR energy-loss rates. In both panels, the DIS contribution increases monotonically with $T_p$, independent of $M_{Z'}$ or $\delta_{\rm DM}$, reflecting the fact that larger $T_p$ implies larger momentum transfer $Q^2$. We also find that a larger mediator mass $M_{Z'}$ causes DIS to dominate at lower values of $T_p$. This can be understood from the propagator factor $(Q^2+M_{Z'}^2)^{-2}$, which controls both the elastic and DIS cross sections. For small $M_{Z'}$, this propagator provides a stronger enhancement in the elastic regime~($Q^2<1\,\mathrm{GeV}^2$) compared to the DIS regime~($Q^2>1\,\mathrm{GeV}^2$). More specifically, once $Q^2 \gtrsim 1\,\mathrm{GeV}^2$~(this will require some minimum value of $T_p$ for fixed $m_{\chi_1}$ and $\delta_{\rm DM}$), DIS invariably dominates the energy loss, even for light mediators. This apparent tension with the corresponding cross-section ratios is resolved by the kinematic advantage of DIS: while elastic scattering transfers only a fraction of the proton energy, DIS can lead to nearly complete energy deposition through hadronic fragmentation. The right panel illustrates that, for fixed DM and mediator masses, increasing the mass splitting suppresses the DIS contribution, which can again be understood from the momentum-transfer relation $Q^2=-\delta_{\rm DM}^2+2\, m_{\chi_1}\, T_{\rm DM}$.
\begin{figure}[]
\centering
\mbox{
\subfigure[]{\includegraphics[width=0.5\textwidth]{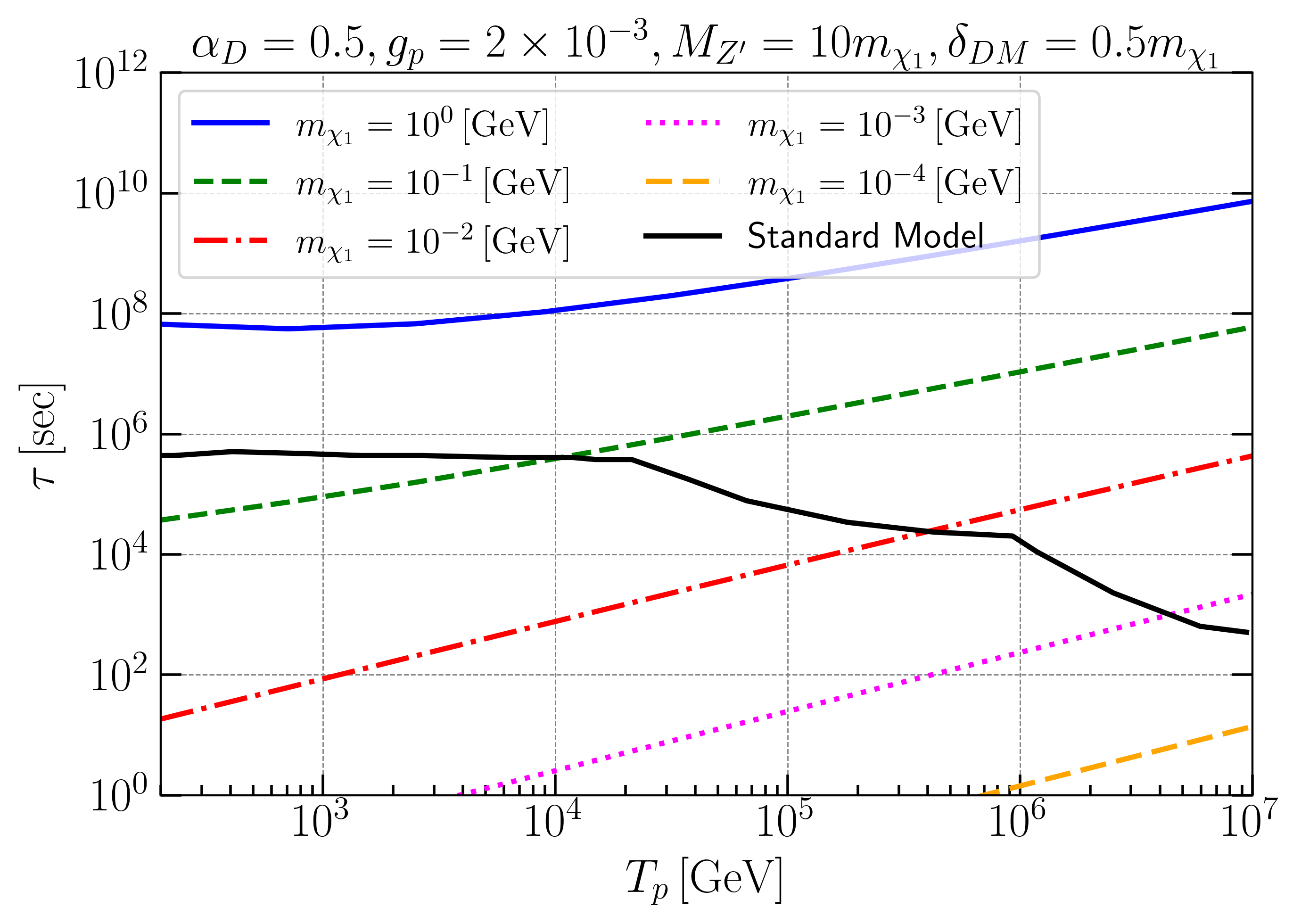}\label{Fig:BP_2a}}
\subfigure[]{\includegraphics[width=0.5\textwidth]{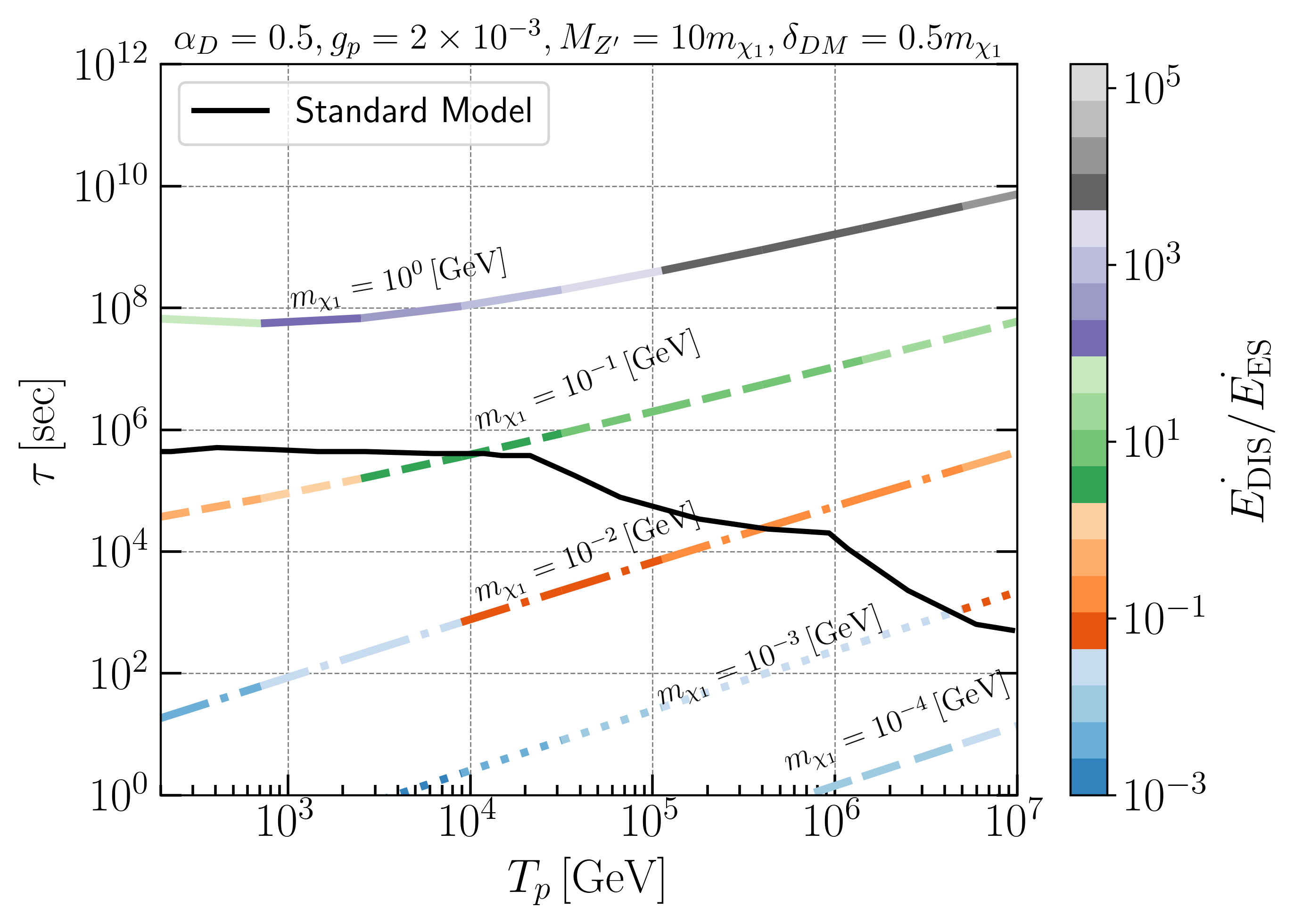}\label{Fig:BP_2b}}
}
\caption{Cooling timescales from CR protons scattering with DM, in terms of proton energy
$T_p$, compared with those from SM processes for different DM masses. We fix the mass splitting between the two DM states to be $\delta_{\rm DM}=0.5\, m_{\chi_1}$ and choose the mediator mass as $M_{Z'} = 10\,m_{\chi_1}$. The right panel is identical to the left panel, with the color bar indicating the DIS contribution.}
\label{fig:cooling}
\end{figure}
\par Fig.~\ref{fig:cooling} shows the dependence of the cooling timescale on the DM mass with the mediator mass and mass splitting fixed to $M_{Z'}=10 m_{\chi_1}$ and $\delta_{\rm DM}=0.5 m_{\chi_1}$. The SM cooling timescale is shown as the solid black line for reference. As $m_{\chi_1}$ increases, the DM number density decreases at fixed average density $\langle \rho_{\chi_1} \rangle$, resulting in longer cooling timescales or equivalently smaller interaction rates. Consequently, lighter DM masses yield stronger constraints. In particular, for $m_{\chi_1}\sim 10^{-2}\,\mathrm{GeV}$, the DM-induced cooling can dominate over the SM cooling channels for $T_p \lesssim 10^6\,\mathrm{GeV}$. The timescales also increase with $T_p$ since scattering becomes inefficient at large-momenta transfer. For low DM masses, the momentum transfer remains small and the DIS contribution to the energy-loss rate is weak, so elastic scattering dominates.  In contrast, increasing $m_\chi$ leads to larger momentum transfer, which enhances the DIS contribution relative to the elastic channel. Consequently, the ratio $\dot{E}_{\rm DIS}/\dot{E}_{\rm ES}$ increases systematically with DM mass, as indicated by the color bar in the right panel of Fig.~\ref{fig:cooling}.
\begin{figure}[]
\centering
\mbox{
\subfigure[]{\includegraphics[width=0.5\textwidth]{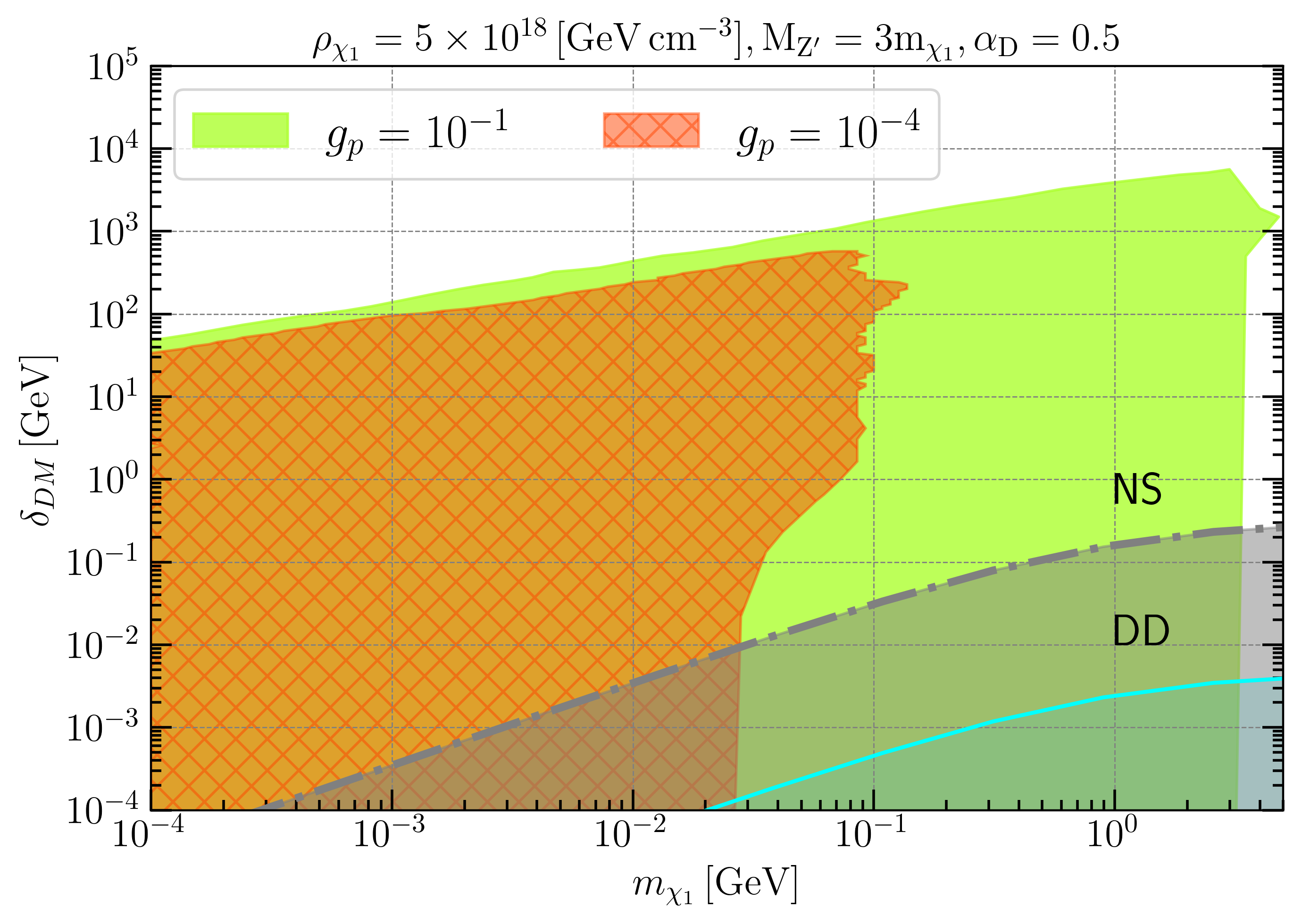}\label{Fig:BP_3a}}
\subfigure[]{\includegraphics[width=0.5\textwidth]{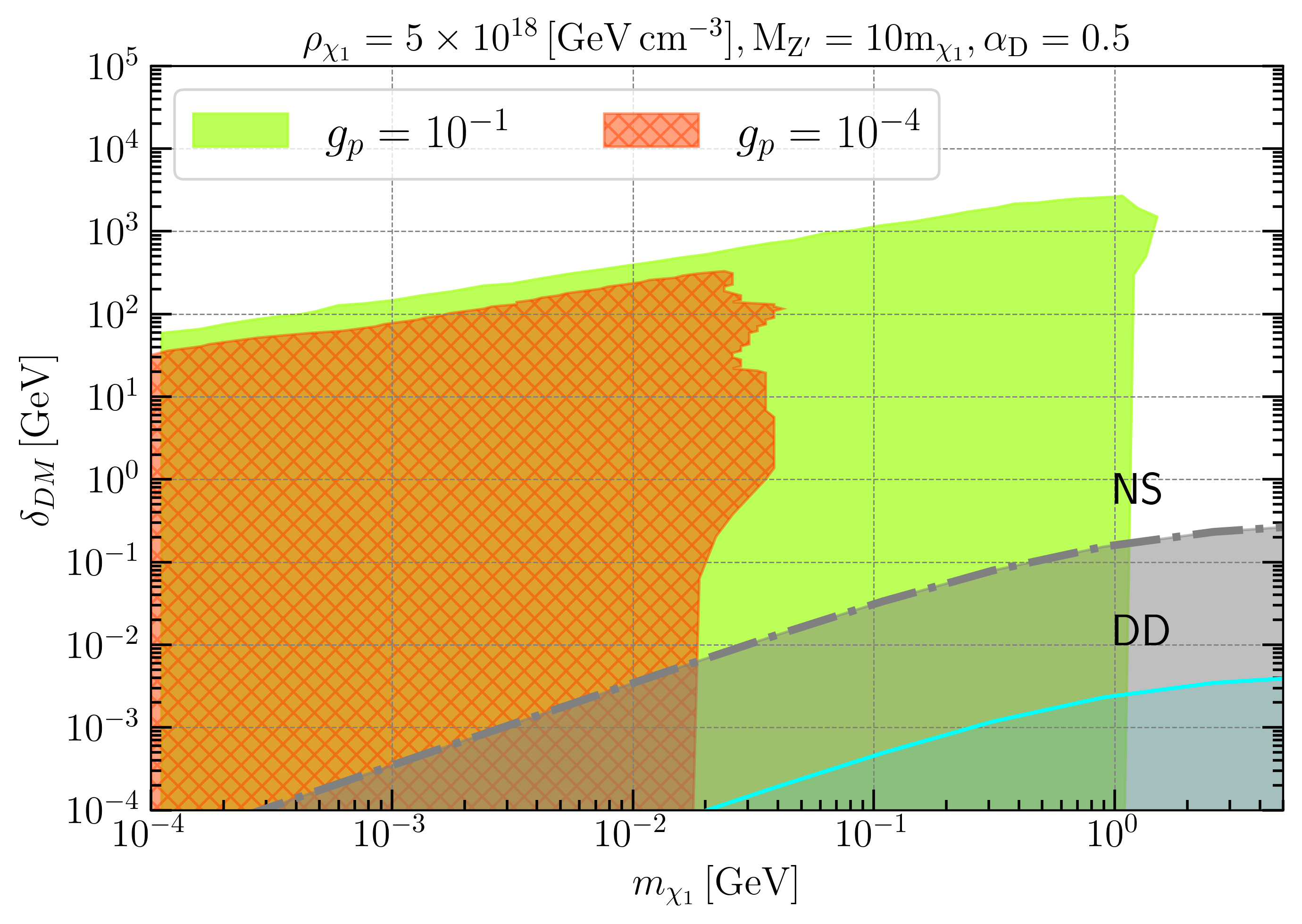}\label{Fig:BP_3b}}
}
\caption{Sensitivity on mass splitting $\delta_{\rm DM}$ as a function of DM mass $m_{\chi_1}$ for two benchmark of proton-DM coupling $g_p=0.1$~(green region) and $10^{-4}$~(orange region). Left and right panel stand for the mediator to DM mass ratios $M_{Z'}/m_{\chi_1}=3$ and $M_{Z'}/m_{\chi_1}=10$, respectively. For comparison, complementary constraints from neutron star (NS) and direct detection (DD) experiments are also shown as gray and cyan shaded regions, respectively.}
\label{Fig:BP_Plot3}
\end{figure}
\par In light of the above discussion, we see that for certain regions of the iDM parameter space, the cooling timescales can become comparable to, or even shorter than, the SM cooling timescale at the relevant energies, in conflict with observations. We also find that the inclusion of DIS systematically reduces the cooling timescales across the entire kinetic-energy range. The iDM model has five free parameters: the DM mass $m_{\chi_1}$, the mass splitting $\delta_{\rm DM}$, mediator mass $M_{Z'}$, and the mediator coupling to the proton $g_p$ and the dark sector $\alpha_{D}=g_\chi^2/4\pi$, which we fix to $\alpha_D=0.5$. We therefore derive constraints or sensitivities in this four-dimensional parameter space, focusing in particular on how large a mass splitting can be probed for sub-GeV DM, which is typically only weakly constrained by direct-detection experiments. 

Fig.~\ref{Fig:BP_Plot3} presents the sensitivity in the $m_{\chi_1}$--$\delta_{\rm DM}$ plane for two representative choices of the mediator-to-DM mass ratio, $M_{Z'}/m_{\chi_1}=3$ (left panel) and $M_{Z'}/m_{\chi_1}=10$ (right panel), and two benchmark couplings $g_p=10^{-1}$ (green) and $g_p=10^{-4}$ (orange). For fixed DM mass, the sensitivity remains strong only up to a certain value of $\delta_{\rm DM}$, beyond which it progressively weakens. This behavior has two origins. First, as $\delta_{\rm DM}$ increases, $Q^2$ decreases, which reduces the suppression of the differential cross section as seen in Eqs.~\ref{eq:diff_elastic} and \ref{eq:diff_DIS}. Second, cooling is kinematically allowed only when
$ s > \left(m_{p} + m_{\chi_1} + \delta_{\rm DM}\right)^2$.
Consequently, the bounds become weaker for $m_{\chi_1} \lesssim
\frac{\delta_{\rm DM}^2 + 2 m_{p}\delta_{\rm DM}}
{2T_{p,\text{min}}-2\delta_{\rm DM}}$
and disappear entirely when $m_{\chi_1} \lesssim
\frac{\delta_{\rm DM}^2 + 2 m_{p}\delta_{\rm DM}}
{2T_{p,\text{max}}-2\delta_{\rm DM}}$, where $T_{p,\text{min}}$ and $T_{p,\text{max}}$ denote the minimum and maximum kinetic energies of the proton considered in the cooling analysis. As expected, larger DM-proton coupling allows a large region of the  $m_{\chi_1}-\delta_{\rm DM}$ plane to be probed, since the corresponding cross-section is larger and the cooling time scale shorter.  Although the DIS contribution can be significant for a heavier mediator mass, a comparison of the two panels shows that a heavier mediator allows sensitivity to smaller DM masses. 
\par The iDM has been investigated in various astrophysical environments, such as neutron stars (NS), as well as in direct detection (DD) experiments. Near a NS, gravitational acceleration can boost $\chi_1$ to $v_{\rm rel}\approx 0.8c$, enabling efficient energy deposition through scattering~\cite{Bell:2020jou,Bell:2018pkk,Alvarez:2023fjj,Roy:2026sek}. Consequently, Cold NSs with surface temperatures of order $2000\,\rm K$ may reveal kinetic heating induced by the deposition of kinetic energy from $\chi_1$ in scattering events, potentially making this scenario accessible to future infrared telescopes~\footnote{For the NS analysis, we assume that $\chi_1$ is in the optically thick regime, such that the capture efficiency is saturated. A detailed study of the resulting observational signatures and detection prospects is beyond the scope of the present work and is left for future investigation.}. In contrast, the boosted DM scenario offers promising detection prospects in Earth-based experiments, where $\chi_1$ with velocities $v_{\rm rel}\approx \mathcal{O}(0.1c)$ can induce recoil energies above detector thresholds~\cite{Giudice:2017zke,Bringmann:2018cvk,Bell:2021xff}. Based on the kinematic requirement for DM up-scattering, we define the upper limit on $\delta_{\rm DM}$ by imposing the following condition:
\begin{align}
   \delta_{\rm DM} \leq \left(\frac{2m_{\chi_1}m_{\rm{SM}}}{\sqrt{1-v^2_{\rm rel}}}+m^2_{\chi_1}+m^2_{\rm{SM}}\right)^{1/2}-m_{\chi_1}-m_{\rm{SM}},
\end{align}
where $m_{\rm SM}$ is the mass of the Standard Model (SM) target and $v_{\rm rel}$ is the relative velocity measured in the rest frame of SM target. Correspondingly, in Fig.~\ref{Fig:BP_Plot3}, we show the DM sensitivity reach for NS (gray shaded region) and DD experiments (cyan shaded region). We see that in the regime of large mass splittings, above a few GeV, conventional direct-detection experiments and astrophysical probes such as neutron stars offer little prospect for discovering iDM. Hence, we conclude that CR cooling in AGN provides sensitivity to inelastic dark matter mass splittings that are orders of magnitude larger than those accessible in conventional searches, reaching well above the TeV scale for DM mass $m_{\chi_1}\sim \mathcal{O}(0.1)$~GeV.
\section{Conclusion}
In this work, we investigated the sensitivity of CR cooling in the AGN NGC 1068 to sub-GeV inelastic dark matter within a minimal vector-portal framework. The dense DM spike surrounding the SMBH can induce significant energy losses of high-energy CR protons through DM upscattering.   Unlike previous analyses that focused mainly on elastic scattering with dipole form factors, we consistently included both elastic and DIS contributions to the cooling rate, demonstrating that DIS becomes dominant at large momentum transfer and substantially enhances the overall cooling effect.

By requiring that the DM-induced cooling timescale not fall below the SM cooling timescale observed in NGC 1068, we derived meaningful constraints on the iDM parameter space. Our results show that AGN cosmic-ray cooling can probe sub-GeV iDM with relatively large mass splittings, extending well beyond the reach of current direct-detection experiments. The sensitivity depends strongly on the mediator mass, the DM–proton coupling, and the structure of the DM spike. Importantly, even when accounting for possible softening of the spike profile due to dynamical evolution effects, the resulting constraints remain competitive and significantly stronger than existing neutron-star and direct-detection bounds over a broad region of parameter space.

Taken together, our study highlights AGNs as powerful astrophysical laboratories for probing non-annihilating sub-GeV inelastic dark matter. The inclusion of DIS contributions is essential for obtaining reliable constraints in the high-energy regime relevant to AGN cosmic rays, and future multimessenger observations offer a promising avenue for further improving the sensitivity to dark-sector interactions.
\acknowledgments{
 The work of S.M. is supported by KIAS
Individual Grants (PG086002) at Korea Institute for Advanced Study. The work of AR is supported by Basic Science Research Program through the National Research Foundation of Korea (NRF) funded by the Ministry of Education through the Center for Quantum Spacetime (CQUeST) of Sogang University (RS-2020-NR049598) and by the Ministry of Science and ICT with grant number RS-2025-24523022.  }
\appendix
\section{Comparison of elastic scattering with dipole form factor, elastic scattering with general form factors and DIS}
\begin{figure}[h]
\centering
\includegraphics[width=0.45\textwidth]{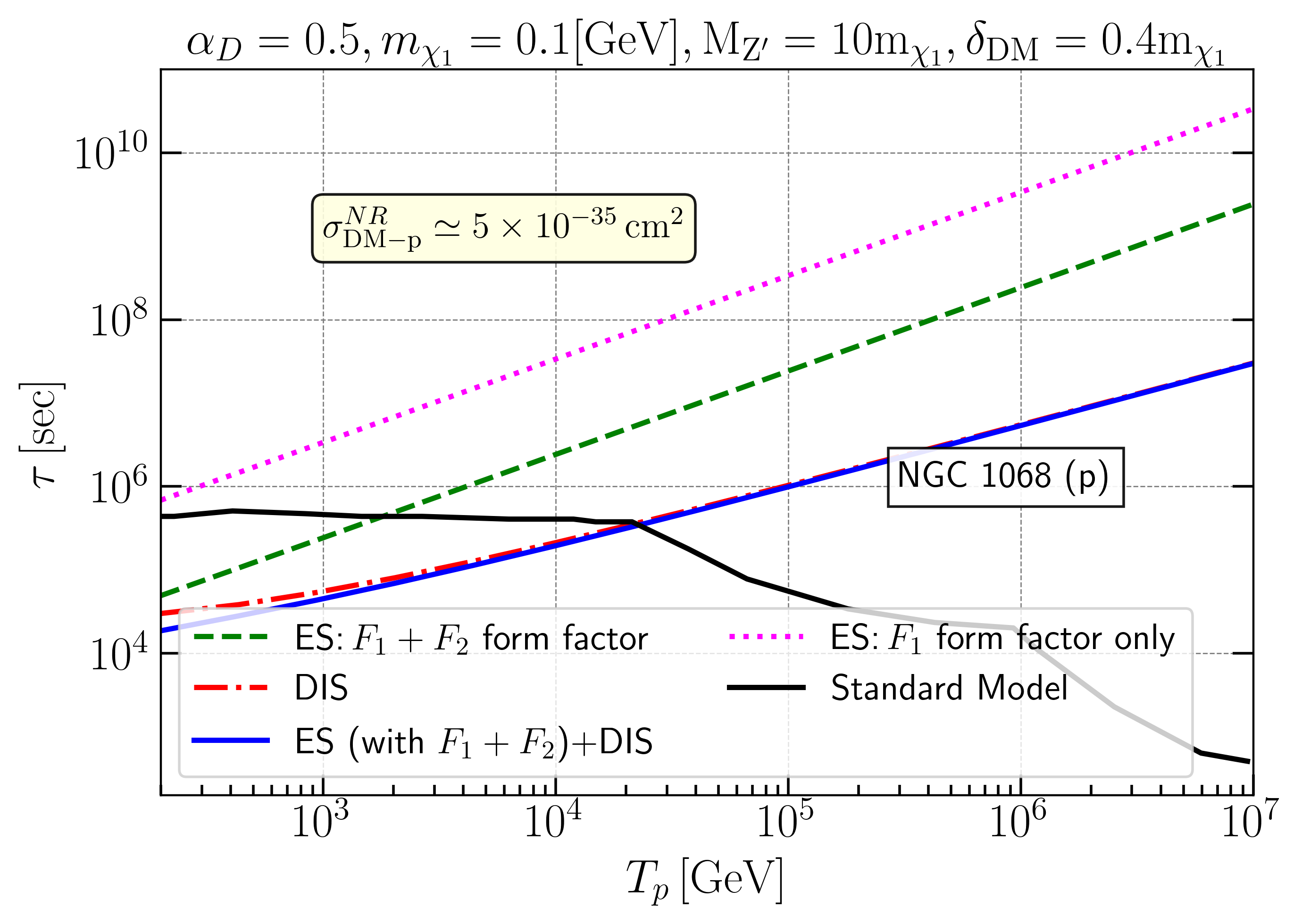}
\includegraphics[width=0.45\textwidth]{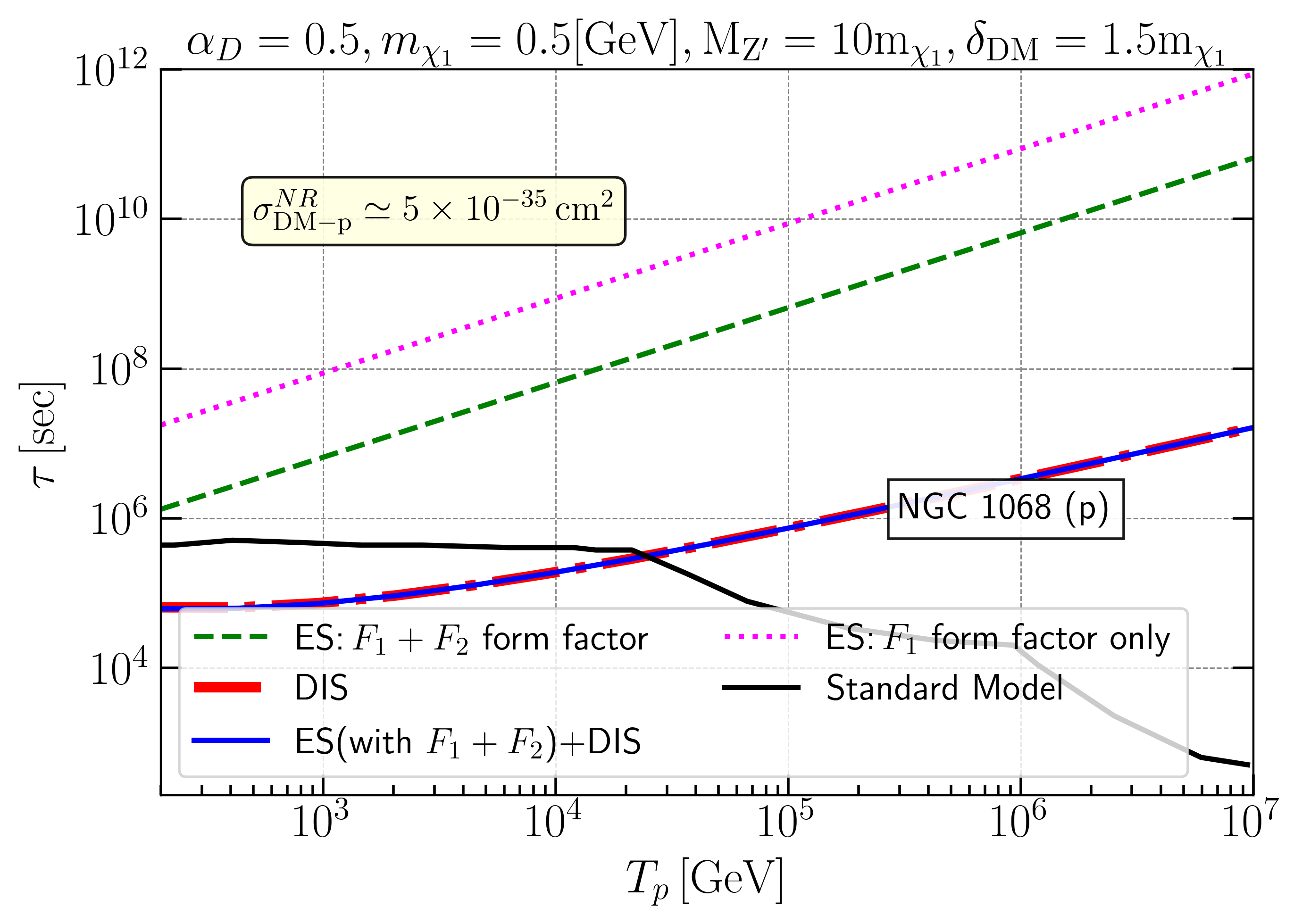}
\caption{Comparison of elastic scattering (ES) with dipole form factor, ES with general form factors and deep inelastic scattering (DIS) contribution to cooling time scale for fixed DM mass and fixed DM mass to mediator ratio.}
\label{fig:Appendix_Plot}
\end{figure}
In Fig.~\ref{fig:Appendix_Plot}, we illustrate the relative importance of the DIS contribution compared to the elastic contribution computed using either the dipole form factor approximation~($F_1(Q^2)=(1+Q^2/\Lambda^2)^{-2}$) or the more general form-factor treatment~($F_1(Q^2),\, F_2(Q^2)\neq 0$). For this analysis, we fix the mediator mass to $M_{Z'}=10\,m_{\chi_1}$, while set the reference DM--proton cross section to $\sigma_{\rm DM-p}^{\rm NR}=5\times10^{-35}\,\text{cm}^2$. The left and right panels differ only in the choice of dark matter mass and mass splitting. We observe that assuming only elastic scattering with the pure dipole form factor~(dotted magenta) leads to significantly larger cooling timescales. Including the full elastic contribution with general form factors~(green dashed) reduces the cooling timescale appreciably. Nevertheless, the DIS contribution, shown by the red dot-dashed curve, provides the dominant contribution to the cooling timescale over the relevant energy range. This highlights the importance of the general form factor, compared to the dipole form factor, in elastic scattering, as well as the overall significance of DIS scattering in cosmic-ray cooling.
\section{DM spike uncertainties}
\begin{figure}[h]
\centering
\includegraphics[width=0.45\textwidth]{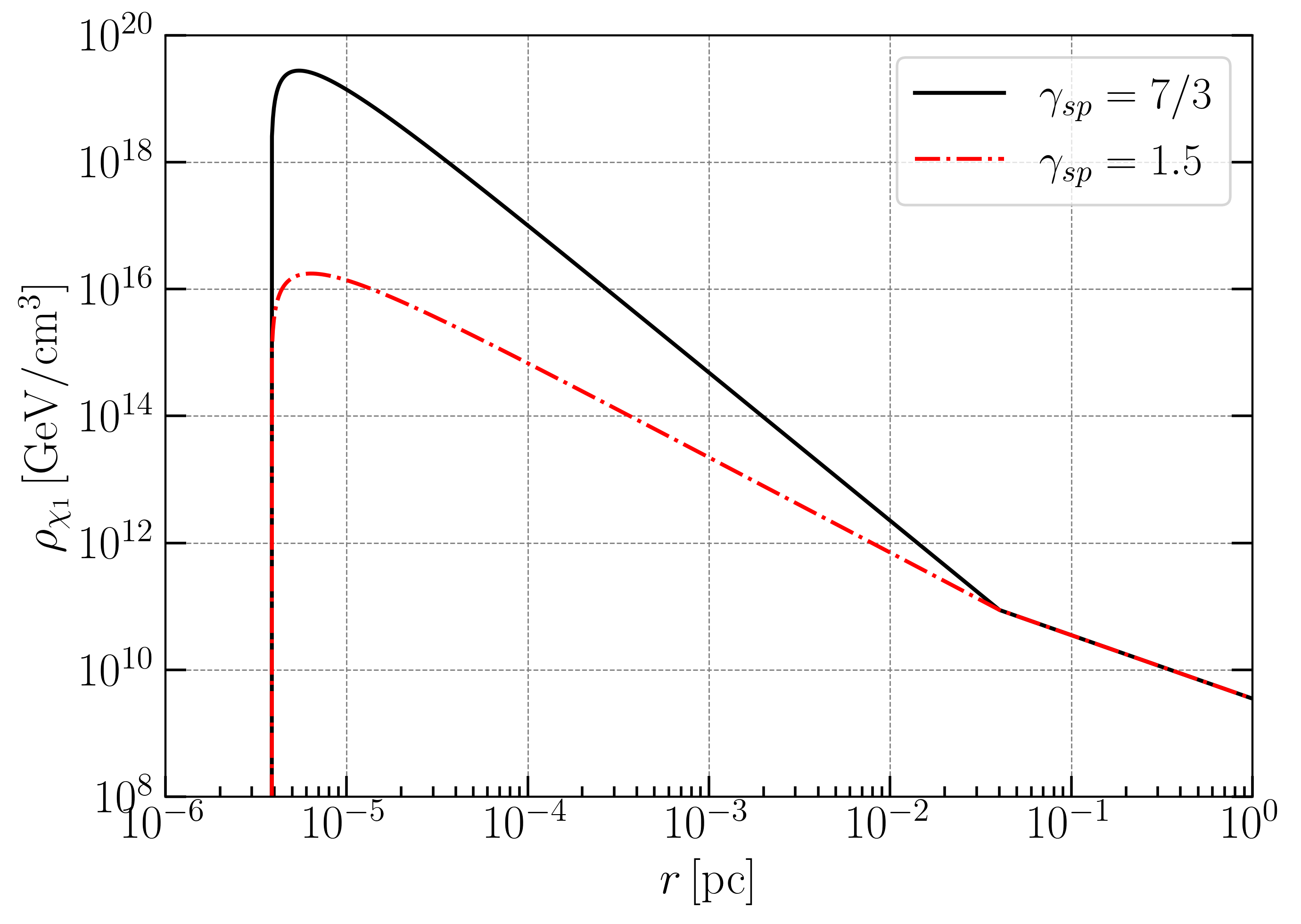}
\includegraphics[width=0.45\textwidth]{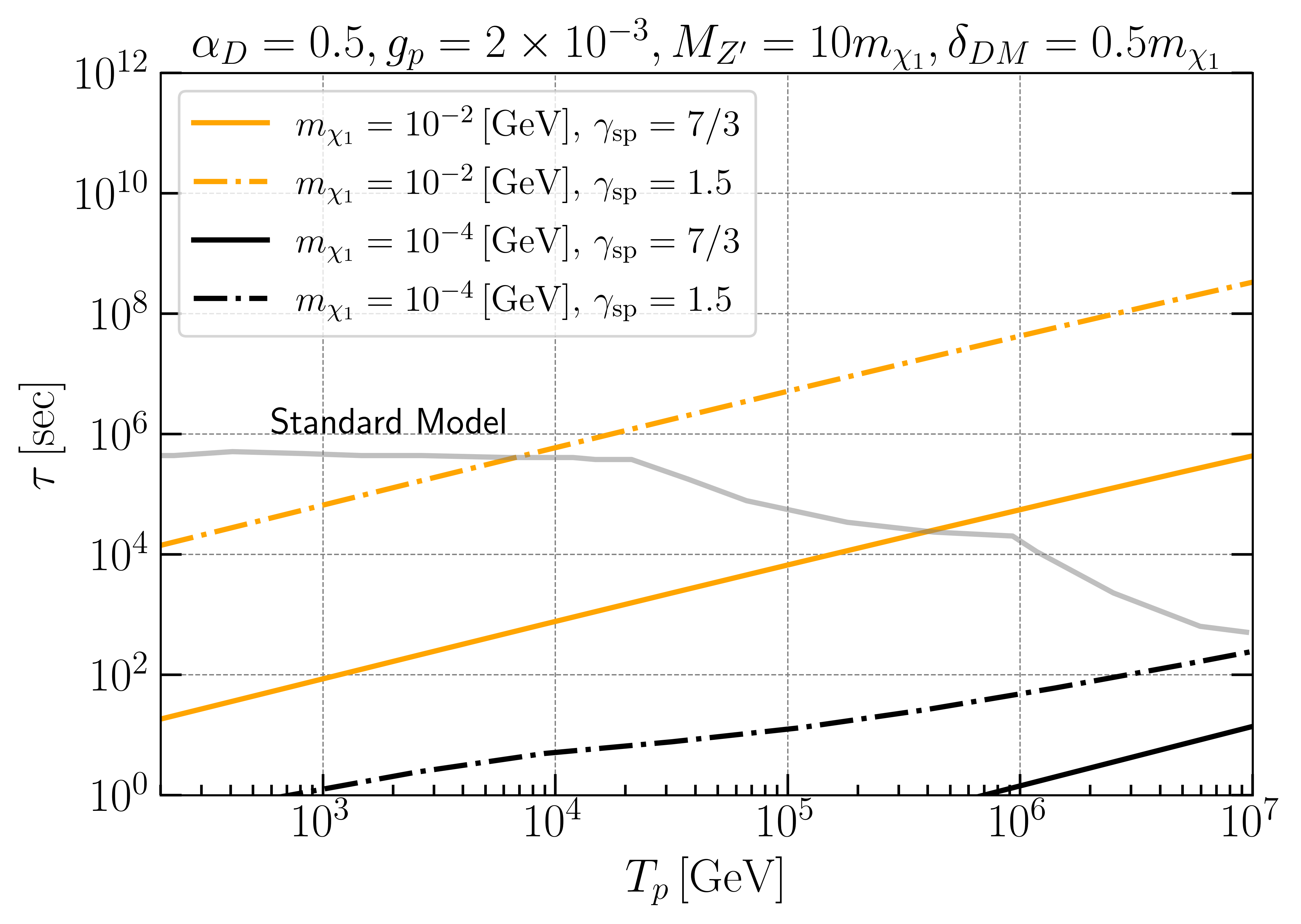}
\caption{Left panel: DM distribution in NGC 1068, for different values of spike index $\gamma_{\rm sp}$. The black line is the standard case with NFW profile and $\gamma=1$. Right panel: dependence of spike index on cooling timescales. Orange and black lines stand for DM mass $m_{\chi_1}=10^{-2}$ GeV and $10^{-4}$ GeV, respectively, whereas the solid and dot-dashed lines stand for spike index $\gamma_{\rm sp}=7/3$ and $\gamma_{\rm sp}=1.5$, respectively.}
\label{fig:Appendix_Plot2}
\end{figure}
In addition to self-annihilations, the steepness of the dark matter spike profile can be reduced by several other mechanisms. In particular, galaxy mergers may lead to the formation of supermassive black hole binaries, which transfer energy to the surrounding dark matter spike and consequently soften the density profile. It has also been suggested that gravitational interactions with baryons can heat the dark matter spike, potentially lowering the spike index to values as small as $\gamma_{\rm sp}=1.5$~\cite{Merritt:2006mt}.
Naturally, such dynamical evolution effects can modify our results. To illustrate their impact, we consider several benchmark choices of $\gamma_{\rm sp}$. In the left panel of Fig.~\ref{fig:Appendix_Plot2}, we show the dark matter profiles in NGC 1068 for different values of $\gamma_{\rm sp}$. Note that the choice $\gamma_{\rm sp}=7/3$ corresponds to an NFW profile with $\gamma=1$. Comparing the black and red curves, one observes that the dark matter spike can be substantially reduced.
In the right panel, we show the corresponding impact on the cooling timescale. For this analysis, we consider two dark matter masses, $m_{\chi_1}=10^{-2}$~GeV (orange curves) and $10^{-4}$~GeV (black curves), together with spike indices $\gamma_{\rm sp}=7/3$ (solid lines) and $\gamma_{\rm sp}=1.5$ (dot-dashed lines). We fix the mediator mass and mass splitting to $M_{Z'}=10,m_{\chi_1}$ and $\delta_{\rm DM}=0.5,m_{\chi_1}$. In both benchmark scenarios, there is a clear difference between the cases with $\gamma_{\rm sp}=7/3$ and $\gamma_{\rm sp}=1.5$. This indicates that the sensitivity obtained in Fig.~\ref{Fig:BP_Plot3} becomes weaker for $\gamma_{\rm sp}=1.5$. More specifically, we find that the sensitivity to the mass splitting is reduced by roughly one to two orders of magnitude, although it still remains stronger than the existing direct-detection and neutron-star constraints.

\bibliographystyle{utphys}
\bibliography{bibliography.bib}
\end{document}